\documentclass{article} 
\usepackage{nips11submit_e,times}

\usepackage{latexsym,bm,amsmath, amssymb, amsthm, bbm, epsf, graphicx, algorithm2e,commath,hyperref}

\title{Capturing the diversity of biological tuning curves using generative adversarial networks}

\author{
Takafumi~Arakaki$^\star$\\
Institute of Neuroscience\\
University of Oregon\\
Eugene, OR 97403 \\
\texttt{tarakaki@uoregon.edu} \\
\And
G.~Barello$^\star$\\
Institute of Neuroscience\\
University of Oregon\\
Eugene, OR 97403 \\
\texttt{gbarello@uoregon.edu} \\
\AND
Yashar~Ahmadian \\
Institute of Neuroscience\\
University of Oregon\\
Eugene, OR 97403 \\
\texttt{yashar@uoregon.edu} \\
\AND
\\
\vspace{-.3cm}
 ${}^\star$ Equal contributions.
}

\date{\normalsize ${}^{\rm @}$  ${}^\star$ Equal
contributions\\ \vspace*{1cm} \today}

\newcommand{\be}{\begin{equation}}
\newcommand{\ee}{\end{equation}}
\newcommand{\bea}{\begin{eqnarray}}
\newcommand{\eea}{\end{eqnarray}}
\newcommand{\lpr}{\left(}
\newcommand{\rpr}{\right)}
\newcommand{\lbr}{\left[}
\newcommand{\rbr}{\right]}

\newcommand{\req}[1]{Eq.~(\ref{#1})}

\newcommand{\eg}{\emph{e.g.},~}
\newcommand{\ie}{\emph{i.e.},~}

\newcommand{\vc}[1]{{\mathbf{#1}}}

\newcommand{\J}{{\mathcal{J}}}

\newcommand{\vr}{{\vc{r}}}

\newcommand{\vx}{{\mathbf{x}}}
\newcommand{\vz}{{\mathbf{z}}}
\newcommand{\G}{{\mathcal{G}}}
\newcommand{\D}{{\mathcal{D}}}
\newcommand{\vtheta}{{\boldsymbol{\theta}}}

\newcommand{\vw}{{\vc{w}}}
\newcommand{\vth}{{\boldsymbol{\theta}}}

\newcommand\ignore[1]{\index{ignore}}

\nipsfinalcopy 

\begin{document}

\maketitle

\begin{abstract}

Tuning curves characterizing the response selectivities of biological neurons often exhibit large degrees of irregularity and diversity across neurons. Theoretical network models that feature heterogeneous cell populations or random connectivity also give rise to diverse tuning curves. However, a general framework for fitting such models to experimentally measured tuning curves is lacking. We address this problem by proposing to view mechanistic network models as generative models whose parameters can be optimized to fit the distribution of experimentally measured tuning curves. 
A major obstacle for fitting such models is that their likelihood function is not explicitly available or is highly intractable to compute. 
Recent advances in machine learning provide 
ways for fitting generative models without the need to evaluate the likelihood and its gradient. Generative Adversarial Networks (GAN) provide one such framework which has been successful in traditional machine learning tasks. 
We apply this approach in two separate experiments, showing how GANs can be used to fit commonly used mechanistic models in theoretical neuroscience to datasets of measured tuning curves. 
This fitting procedure avoids the computationally expensive step of inferring latent variables, \eg~the biophysical parameters of individual cells or the particular realization of the full synaptic connectivity matrix, and directly learns model parameters which characterize the \emph{statistics} of connectivity or of single-cell properties.
Another strength of this approach is that it fits the entire, joint distribution of experimental tuning curves, instead of matching  a few summary statistics picked  \emph{a priori} by the user. More generally, this framework opens the door to fitting theoretically motivated dynamical network models directly to simultaneously or non-simultaneously recorded neural responses.

\end{abstract}

\section{Introduction}
\label{sec-1}

%
Neural responses in many brain areas are tuned to  external parameters such as stimulus- or movement-related features. 
Tuning curves characterize the dependence of neural responses on such parameters, and are a key descriptive tool in neuroscience. 
Experimentally measured tuning curves often exhibit a rich and bewildering diversity across neurons in the same brain area, which complicates simple understanding \cite{Ringach:2002}. 
This complexity has given rise to a tendency towards biased selections of minorities of cells  which exhibit pure selectivites, and have orderly and easily interpretable tuning curves. 
As a result the biological richness and diversity of tuning curves in the full neural population is often artificially reduced or ignored. 
On the theoretical side too, many network models feature  homogeneous populations of cells with the same cellular parameters and with  regular synaptic connectivity patterns. Neural  tuning curves in such models will naturally be regular and have identical shapes.

New theoretical advances, however, have highlighted the computational importance of diverse tuning and mixed selectivity, as observed in biological systems \cite{Rigotti:2013, Barak:2013}.
Furthermore, diversity and heterogeneity can  be produced in mechanistic network models which either include  cell populations with heterogeneous single-cell parameters (see \eg \cite{Persi:2011}), or connectivity that is partly random and irregular despite having regular \emph{statistical} structure (see, \eg \cite{Roxin:2011,Litwin-Kumar:2012, Barak:2013,  Rubin:2015, Ahmadian:2015, Lalazar:2016}). 
However, a general effective methodology for fitting such models to experimental data, such as heterogeneous samples of biological tuning curves is lacking.

A related central problem in  neural data analysis is that of inferring functional and synaptic connectivity from neural responses and correlations. A rich literature has addressed this problem \cite{Schneidman:2006, Shlens:2006, Pillow:2008, Tang:2008, Shlens:2009, Yatsenko:2015}. However, we see two shortcomings in previous approaches. First, most methods are based on forward models that are primarily inspired by their ease of optimization and fitting to  data, rather than by theoretical or biological principles. Second, in the vast majority of approaches, the outcome is the estimate of the particular connectivity matrix between the particular subset of neurons sampled and simultaneously recorded in particular animals \cite{Schneidman:2006, Shlens:2006, Pillow:2008, Tang:2008, Shlens:2009, Yatsenko:2015}. Post-hoc analyses may then be applied  to characterize various statistical properties and  regularities of connectivity \cite{Shlens:2006, Yatsenko:2015}. The latter are what is truly of interest, as they generalize beyond the particular recorded sample.  Examples of such statistical regularities are the dependence of  connection probability between neurons on their physical distance \cite{Perin:2011} or preferred stimulus features \cite{Ko:2011}. Another example is the degree to which neuron pairs tend to be connected bidirectionally beyond chance   \cite{Song:2005}.
A methodology for inferring or constraining such statistics directly from simultaneously or non-simultaneously recorded neural responses is lacking. 
%

Here we propose a methodology that is able to fit theoretically motivated network models to (simultaneously or non-simultaneously) recorded neural responses, by directly optimizing parameters characterizing the \emph{statistics} of connectivity or of single-cell properties. Conceptually, we propose to view  network models with heterogeneity and random connectivity as generative models for the observed neural data, \eg~a model that generates diverse tuning curves and hence implicitly models their (high-dimensional) statistical distribution. The generative model is determined by the set of network parameters which specify the distribution of mechanistic circuit variables like synaptic connectivity matrix or single-cell biophysical properties. In this picture, the particular realization of the connectivity matrix or of biological properties of particular  neurons are viewed as latent variables. 
Traditional, likelihood-based approaches such as expectation maximization or related approaches need to optimize or marginalize out (\eg using variational or Monte Carlo sampling methods) such latent variables conditioned on the particular observed data sample. Such high-dimensional optimizations or integrations are computationally very expensive and often intractable. 

Alternatively, one could fit theoretical circuit models by approaches similar to moment-matching, or its Bayesian counterpart, Approximate Bayesian Computation \cite{Marin:2011,Beaumont:2010}. In such approaches, one \emph{a priori} comes up with a few summary statistics, perhaps motivated on theoretical grounds, which  characterize the data objects (\eg tuning curves). Then one tunes (or in the Bayesian case, samples) the model parameters (but not latent variables) so that the few selected summary statistics are approximately matched between  generated tuning curve samples and experimental ones \cite{Lalazar:2016}. 
This approach will, however, generally be biased by the \emph{a priori} choice of  statistics to be fit. Furthermore, when model parameters are more than a few (such that grid search becomes impractical), and the chosen statistics are complex and possibly non-differentiable, the fitting procedure will become impractical.  By contrast, in the approach we propose here, the fitting of the high-dimensional data distribution can in principle be done in a much more unbiased manner,\footnote{As long as the discriminator network (see Sec.~\ref{subsec-21}) used in the adversarial training is sufficiently complex.} and without the need to \emph{a priori} choose a few summary statistics.

A suite of new methods have recently been developed in machine learning for fitting \emph{implicit} generative models \cite{Kingma:2013,Goodfellow:2014,Tran:2017}, \ie~generative models for which a closed or tractable expression for the likelihood or its gradient is not available. Here, we will use one of these methods, namely Generative Adversarial Networks (GANs) \cite{Goodfellow:2014,WGAN,ImprovedWGAN}. However, other approaches are worth noting, including variational autoencoders \cite{Kingma:2013,Doersch:2016}, and hierarchical and deep implicit models \cite{Tran:2017}. These other methods are potentially as powerful in their application to neuroscience. Additionally, there is recent progress in unifying these approaches \cite{Mescheder:2017,hu_unifying_2017,rosca_variational_2017} which may clarify future applications of these techniques. This report is organized as follows. In section \ref{sec-2},  we introduce the general GAN framework, and  the two circuit models adopted from the theoretical neuroscience literature that we use in our experiments. In section \ref{sec-3}, we present the results of our GAN-based training of these two models. Finally, in sections \ref{sec-4} and \ref{sec-5} we discuss areas for improvement of our proposed methodology, and other potential future applications of it.

\section{Methods}
\label{sec-2}

\subsection{Generative Adversarial Networks}
\label{subsec-21}

Generative Adversarial Networks (GANs) are a recently developed approach to fitting generative models to unlabeled data, such as natural images \cite{Goodfellow:2014,DBLP:journals/corr/Goodfellow17,Radford:2015}.  The GAN approach is powerful because it is applicable to models for which evaluating the likelihood or its gradient are intractable; all that is required is a generative process that, given a random seed, generates a sample data object. In particular, the GAN approach avoids the computationally costly step of inference that is required in, \eg~the expectation maximization algorithm.

In a GAN there are two networks. The first network, the ``generator'', implements the generative model, and produces data samples. The generator is defined by a function $\G$ which generates a data sample $\G(\vz)$ given some sample $\vz$ from a fixed, standard distribution. The other network is the ``discriminator'', given by a function $\D$ acting on the data space, which is trained to distinguish between the real data and the samples generated by $\G$. The generator is trained to fool the discriminator. When both $\D$ and $\G$ are differentiable functions, training can be done in a gradient-based way. If the discriminator network is sufficiently powerful, the only way for the generator to fool it is to effectively generate samples that effectively have the same distribution as that of the real data. 

GANs are often difficult to optimize and many tricks and techniques have been developed to make learning more robust. In this work we employ the Wasserstein GAN (WGAN) approach which has shown promise in overcoming some of the short-comings of generators learned in the traditional GAN approach \cite{WGAN,ImprovedWGAN}. We note, however, that traditional GAN's could also be used for the types of application we have in mind.\footnote{In its original formulation, unlike in WGAN, the GAN training is equivalent to a zero-sum game with a loss-function equal to the cross-entropy loss (a measure of classification error) for the discriminator. The discriminator parameters are changed to minimize this loss while the generator parameters are changed to maximize it, leading to a minimax or zero-sum game \cite{DBLP:journals/corr/Goodfellow17}.}
 The insight of the WGAN approach is to add a term to the discriminator's loss which penalizes too small or too large gradients with respect to the inputs, the effect of which is to avoid the vanishing and exploding gradient problem which regular GANs suffer from.\footnote{In the WGAN approach, the $\D$ network is not necessarily a classifier, but we will nevertheless keep referring to it as the ``discriminator".}
This approach is mathematically motivated by minimizing the Wasserstein distance (also known an the earth mover's distance) between the data distribution and the generative model's distribution \cite{WGAN}. The Wasserstein distance provides a measure of similarity between  distributions which exploits metric structure in the data space (by contrast, maximum likelihood fitting minimizes the Kullback-Leibler divergence between the data and model distributions, which is a purely information theoretic measure, blind to metric structure  in data space).

Denoting the data samples by $\vx$, with $\epsilon$ a uniformly distributed random variable between $0$ and $1$, the modified WGAN losses \cite{ImprovedWGAN} for $\D$ and $\G$ are given by
\begin{align}
\text{Loss}_{\mathcal{D}} &= \langle \mathcal{D}(\mathcal{G}(\mathbf{z}))\rangle_{\mathbf{z}} - \langle \mathcal{D}(\mathbf{x})\rangle_{\mathbf{x}} + \lambda \langle \left(\|\nabla \mathcal{D}(\epsilon \mathbf{x} + (1 - \epsilon) \mathcal{G}(\mathbf{z}))\|_2 - 1 \right)^{2}\rangle_{\mathbf{z},\mathbf{x},\epsilon} \label{eq:loss-d}\\
\text{Loss}_{\mathcal{G}} &= - \langle \mathcal{D}(\mathcal{G}(\mathbf{z}))\rangle_{\mathbf{z}} \label{eq:loss-g},
\end{align}
respectively. The average over $\vx$ denotes the empirical average over the training examples, while $\vz$ and $\epsilon$ are averaged over their fixed, standard distributions.  When necessary to make the dependence of discriminator and generator functions on their learned parameters $\vw$ and $\vth$ explicit, we denote them by $\D_{\vw}$ and $\G_{\vth}$, respectively.  A stochastic gradient descent algorithm on these loss functions is shown in panel \ref{alg-1}.

\begin{algorithm}[H]\label{alg-1}
    \KwIn{data distribution \(\mathbb P_r\), the gradient penalty coefficient  \(\lambda\), the number of critic iterations per generator iteration \(n_{\text{critic}}\) , the batch size \(m\), Adam hyperparameters \(\alpha_{\text{critic}}, \alpha_{\text{generator}}, \beta_1, \beta_2\) and the initial critic \(\mathbf{w}_0\) and generator \(\vtheta_0\) parameters.}

    \(\vtheta \leftarrow \vtheta_0\)\;
    \(\mathbf{w} \leftarrow \mathbf{w}_0\)\;
    \While{\(\vtheta\) has not converged}{
        \Repeat{critic is updated \(n_{\text{critic}}\) times}{
            \For{\(i = 1, ..., m\)}{
                Sample real data \(\mathbf{x} \sim \mathbb P_r\),
                latent variable \(\mathbf{z} \sim p(\mathbf{z})\),
                a random number \(\epsilon \sim U[0, 1]\).\;
                \(\tilde{\mathbf{x}} \leftarrow \mathcal G_\vtheta(\mathbf{z})\)\;
                \(\hat{\mathbf{x}} \leftarrow \epsilon \mathbf{x} + (1 - \epsilon) \tilde{\mathbf{x}}\)\;
                \(\text{Loss}_{\mathcal{D}}^{(i)} \leftarrow \mathcal{D}_{\mathbf{w}}(\tilde{\mathbf{x}}) - \mathcal{D}_{\mathbf{w}}(\mathbf{x}) + \lambda(\|\nabla_{\hat{\mathbf{x}}} \mathcal{D}_{\mathbf{w}}(\hat{\mathbf{x}})\|_2 - 1)^2\)\;
            }
            \(\mathbf{w} \leftarrow \text{Adam}(\nabla_{\mathbf{w}} \frac{1}{m} \sum_{i=1}^m \text{Loss}_{\mathcal{D}}^{(i)}, \mathbf{w}, \alpha_{\text{critic}}, \beta_1, \beta_2)\)\;
        }
        Sample a batch of latent variables \((\mathbf{z}^{(i)})_{i=1}^m \sim p(\mathbf{z})\).\;
        \(\vtheta \leftarrow \text{Adam}(- \nabla_\vtheta \left( \frac{1}{m} \sum_{i=1}^m \mathcal{D}_{\mathbf{w}}(\mathcal{G}_\vtheta(\mathbf{z}^{(i)})) + \text{Penalty}_{\mathcal{G}} \right), \vtheta, \alpha_{\text{generator}}, \beta_1, \beta_2)\)\;
    }
    \caption{Improved WGAN algorithm \cite{ImprovedWGAN}. Most of the hyper-parameters are taken from \cite{ImprovedWGAN}: \(n_{\text{critic}} = 5, \beta_1=0.5, \beta_2=0.9\). We use \(\alpha_{\text{critic}}=.001, \alpha_{\text{generator}}=.001, m=30, \text{Penalty}_{\mathcal{G}} = 0\) for the feedforward model and \(\alpha_{\text{critic}}=0.02, \alpha_{\text{generator}}=0.01, m=15, \text{Penalty}_{\mathcal{G}} =\) Eq. \ref{eq:ssn_penalty} for the SSN.}
\end{algorithm}

\subsection{Generative model examples from theoretical neuroscience}
\label{subsec-examplemodels}

In this article, we consider two network models, developed in theoretical neuroscience, as basic examples of generator networks, to illustrate the general GAN-based appraoch we are proposing for fitting heterogeneous network models to experimental data. In both examples the experimental data take the form of a collection of tuning curves for different neurons from some brain area (tuning curve is the function describing the dependence of a neuron's trial-averaged response on one or several stimulus or behavioral parameters). {But our proposed methodology, based on viewing such models as implicit generative models, is more general and can (with small modifications) also be applied to fit network models to data beyond tuning curves (time-series data from simultaneously recorded neurons in different experimental conditions constitute another important example).} Both example networks considered here are scientifically grounded, previously published models of biological neural networks and neural response tunings. In particular, their parameters in principle correspond to physiological and anatomical parameters of biological neural circuits. This is in contrast to the usual case in the GAN literature where the structure of the generator (\eg a deconvolutional deep feedforward network, generating natural images) has no direct mechanistic interpretation.
    
In both examples, the random variability occurs in the network structure (connection strengths). By contrast, the external inputs to the network's neurons, which represent experimental stimuli or conditions used in training data, range over a fixed set of possibilities in all random realizations of the network. To avoid confusion, we will refer to the random variables (denoted by $\vz$ as in Sec.~\ref{subsec-21})  determining the network structure as ``random inputs'', and will refer to the neurons' external inputs as ``stimuli''. We index stimuli ranging over the set of $S$ possibilities used in training data by $s \in \{1,\ldots,S\}$. Note that a trained network model can nevertheless be applied to stimuli other than those it is trained on.
Let $G(s,\vz)$ denote the response of a selected neuron (chosen in advance) in a specific realization of the network (determined by $\vz$) to stimulus $s$.  Throughout we will use $\mathcal{G}(\mathbf{z})$ to denote the $S$-dimensional vector $(G(s,\vz))_{s=1}^S$. Thus $\G(\vz)$ is nothing but the discretized tuning curve of the selected  neuron, \ie the set of its responses to each of several stimuli differing in one or more stimulus parameters.

\subsubsection{Feedforward Model}
\label{sec:lalazar}

As our first example we take a feedforward model of primary motor cortex (M1) tuning curves proposed by \cite{Lalazar:2016}.\footnote{Reference \cite{Lalazar:2016} introduced their model in two versions, a basic one, and an ``enhanced" version. We have used their enhanced version, with small modifications which allow our approach to be used,  as our first example.}
 This model was proposed by \cite{Lalazar:2016} as a  plausible mechanism for generating the significant complex nonlinear deviations of observed M1 tuning curves from the classical linear model \cite{Georgopoulos:1982}. 
The authors fit that model to a collection of experimentally measured M1 tuning curves by matching two specific statistics which were of interest for theoretical reasons. These tuning curves describe the responses of a given neuron in 54 experimental conditions corresponding to the monkey holding its hand, either in supination or pronation, in one location out of a $3\times 3 \times 3$ cubic grid of spatial locations. In our (numerical) experiment we only used the pronation data, and we will refer to the 27 conditions as ``stimuli", and denote them by $s \in \{1,\ldots,27\}$, while denoting the  3D hand location in condition $s$ by $\vx_s$. 

The model is a two-layer feedforward network, with  an input layer, putatively corresponding to the parietal reach area or to premotor cortex, and an output layer modeling M1 (see Fig. \ref{SSNgraphics}). The activations of the input layer neurons are determined by Gaussian receptive fields on three dimensional space. Across the input layer, the Gaussian receptive fields are centered on a regular cubic  grid that extends three times beyond the $3\times 3\times 3$ stimulus grid along each dimension. Whereas Lalazar et al.~used a network with $100$ input-layer receptive fields along each axis, we reduced this number to $40$ to allow faster computations; however, changing this resolution beyond 40 only weakly affects the results. The receptive field widths were random across the input layer, and were sampled iid from the uniform distribution on the range $[\sigma_{l},\sigma_{l} + \delta\sigma]$.
The feedforward connections from the input to output layer are sparse, with a sparsity of $1\%$. In our implementation of this model, the strength of the nonzero  connections were sampled iid from the uniform distribution on the range $[0,J]$. The response of an output layer neuron is given by a rectified linear response function with a threshold. In the original model of \cite{Lalazar:2016} the thresholds were chosen separately for each model neuron such that the coding level of its response (defined  as the fraction of stimuli to which a neuron responds with a rate significantly different from zero)  matched that of a randomly selected data neuron. In order to have all structural variability in the model in a differentiable form, we instead sampled the threshold uniformly and independently from the range $[\phi_{l},\phi_{l} + \delta\phi]$.
We found that with this addition the model was in fact able to match the distribution of neural coding levels in the dataset of \cite{Lalazar:2016}.
%
\begin{table}
\begin{center}
\begin{tabular}{c|c}
Parameter&Description\\
\hline
$\sigma_{l}$&Lower bound of receptive field size range\\
$\delta\sigma$&Width of receptive field size range\\
$J$&Scale of connection strengths\\
$\phi_l$&Lower bound of threshold range\\
$\delta\phi$&Width of threshold range
\end{tabular}
\end{center}
\caption{The parameters of the feedforward model, which are fit to tuning curve data.}
\label{FFparam}
\end{table}

The collection of random inputs, $\vz$, to the generator function for this model determine all input-layer receptive field sizes, individual feedforward connection strengths, and all output-layer neural thresholds for a particular network realization. The output of the network are responses for a single output layer neuron to the 27 stimuli (since the network is feedforward, with independent connections, one could equivalently sample several neurons from the output layer in each realization). The response $G(s)$ of this network given a stimulus $s$ is calculated by
\be
 G(s;\vz) = \lbr  \sum_{i=1}^{40^3} \J_{i}\,  h_{i}(s) - \phi \rbr_{+}
\ee
where
\bea
&& h_{i}(s) = \frac{1}{Z}\exp\! \left(- \frac{1}{2 \sigma_{i}^{2}}\|\mathbf{x}_s - \bar\vx_i\|^{2}\right) \\
&& \J_i =  J\, z^J_i\, M_i\\
&& \phi = \phi_{l}  + z^{\phi} \delta\phi,  \\
&& \sigma_{i} = \sigma_{l} + z^{\sigma}_i \delta\sigma,\\
&& z^{\phi}, z^J_i, z^{\sigma}_i\,\,  \overset{iid}{\sim}\,\, U[0,1]
\eea
where $Z$ is a normalizing factor such that $\sum_{i} h_{i}(s) = 1$, $\vc{M} = (M_{i})_{i=1}^{40^3}$ is the sparse binary mask with sparsity $1\%$, and $[u]_+ = \max(0,u)$ denotes rectification. The random inputs $\vz$ to this model are  
$ \vz = (z^\phi, (z^{\sigma}_i)_{i=1}^{40^3},(z^J_i)_{i=1}^{40^3},(M_i)_{i=1}^{40^3})$, and there are five trainable parameters $\vth = (\sigma_l,\delta\sigma,J,\phi_l,\delta\phi)$ (listed in table \ref{FFparam}). Crucially, since the network's output, $G(s;\vz)$, is differentiable with respect to each of the trainable parameters, we can use the output gradient with respect to parameters to optimize the latter using any variant of the stochastic gradient descent algorithm.

\subsubsection{Recurrent Model}
\label{subsec:SSNmodel}

The second model we consider is a recurrent model of cortical circuitry, the Stabilized Supralinear Network (SSN), which has found broad success in mechanistically explaining the contextual and attentional modulations of neural responses, across multiple areas of sensory and association cortex \cite{Ahmadian:2013,Rubin:2015, Hennequin:2017}. The SSN is a model of cortical circuitry (within a given cortical area), in which neurons have a supralinear rectified power-law input/output function, $f(u) = k [u]_+^n$ (where  $[u ]_+ = \max(0,u)$, $k>0$, and $n >1$). We consider a topographically organized version of this model, with a one-dimensional topographic map {which could correspond, \eg~to the tonotopic map of the primary auditory cortex (A1), or a one-dimensional reduction of the retinotopic map in primary visual cortex (V1)}. The network is composed of  excitatory ($E$) and inhibitory ($I$) neurons, with one neuron of each type at each  topographic spatial location. For $N$ topographic locations, the network thus contains $2N$ neurons.  For a network with recurrent connectivity matrix $\mathbf{W}$, the vector of firing rates $\mathbf{r}$ is governed by the differential equation
\begin{equation}
\mathbf{T} \frac{d\mathbf{r}}{dt} = - \mathbf{r} + f\left(\mathbf{W} \mathbf{r} + \mathbf{I}(s)\right),
\qquad\qquad
{f}(\mathbf{u}) \equiv \left( f(u_{i}) \right)_{i=1}^{2N} = \left( k [u_{i}]_+^n \right)_{i=1}^{2N}
\end{equation}
where the diagonal matrix $\mathbf{T} = \text{Diag}\!\left( (\tau_i)_{i=1}^{2N} \right)$ denotes the neural relaxation time constants, and $\vc{I}(s)$ denotes the external or stimulus input in condition $s \in \{1,\ldots,S\}$. Below, for an arbitrary neuron $i$, we denote its type by $\alpha(i)\in \{E,I\}$ and its topographic location by $x_i$. In our SSN example, we let $x_i$'s range from $-4$ to $4$ on a regular grid.
In the SSN example we consider here, we chose to have all random structural variability occur in the connectivity matrix $\vc{W}$. The standard random variables of the GAN formalism, $\vz$, are thus those which determine a concrete realization of $\vc{W}$, and the learned model parameters $\vth$ are those governing the statistics of connectivity (see \req{Wensemble} below).

For large $N$ and with suitable choices of parameters, the SSN will generically reach a stable fixed point \cite{Ahmadian2013}. This fixed point, which we denote by $\hat{\vr}$, represents the steady-state response of cortical neurons to stimuli.  The steady state response for an \emph{a priori} selected neuron, \ie a certain component of $\hat{\vr}$, with index $i_*$, is the output of our SSN, viewed as a generative model in the GAN setting. We set $G(s,\vz) = \hat{r}_{i_*}$ when the SSN stimulus is $I(s)$ so that $\G(\vz) = (G(s,\vz))_{s=1}^S$ is the tuning curve of neuron $i_*$, defined as the set of its steady-state responses in all stimulus conditions.  
   
\begin{figure}
\begin{centering}
\includegraphics[height = 1.7in]{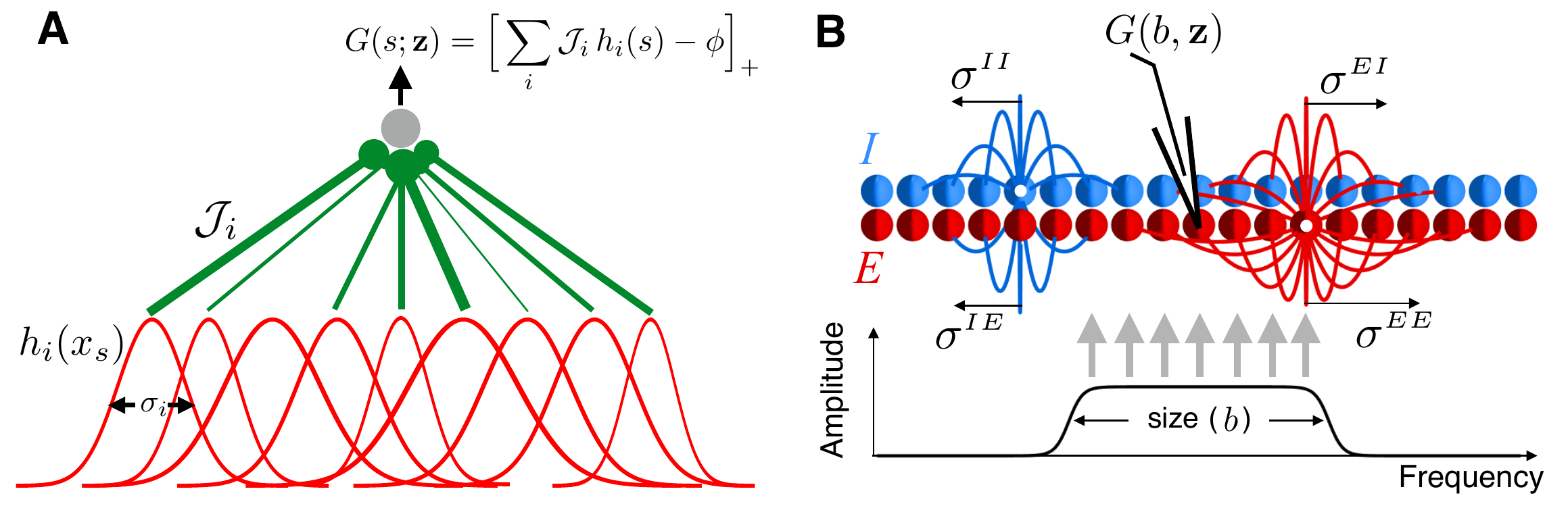}
\caption{Structure of the feedforward (A) and recurrent SSN (B) models used in our  experiments as example generator networks. }
\label{SSNgraphics}
\end{centering}
\end{figure}

In our example, {we simulated the fitting of an SSN model of V1 to datasets of stimulus size tuning curves of V1 neurons} \cite{Rubin:2015}.  We let the stimulus input to neuron $i$ be
\be\label{stimulus}
I_i(s) = A\, \sigma\!\lpr l^{-1}( {{b_s}/{2}+x_i})\rpr
\sigma\!\lpr l^{-1}( {{b_s}/{2}-x_i})\rpr
\ee
where $\sigma(u) = (1+\exp(-u))^{-1}$ is the logistic function, $A$ denotes the stimulus intensity, and $(b_s)_{s=1}^S$ are a set of stimulus sizes. 
Thus, in condition $s$, the stimulus  targets a central band of width $b_s$ centered on the middle of the topographic grid (see Fig.~\ref{SSNgraphics} bottom).  The parameter $l$ determines the width of the smoothed edges of this banded stimulus. This stimulus is a simple and reduced model of visual input to V1 for gratings of diameter $b$.  The tuning curve $\G(\vz)$ is then the stimulus size tuning curve of neuron $i_*$. 
In experiments, typically the grating stimulus used to measure a neuron's size tuning curve is centered on  the neruon's receptive field. To model this stimulus centering, we let the selected neuron, the steady-state responses of which form $\G(\vz)$, be the neuron in the center of the topographic grid, with $x_{i_*}=0$.
\footnote{Note that in general, the tuning curves in random SSN's also show variability across neurons as well as across different realizations of the network for a fixed output neuron. Furthermore, when $N$ is large, one could have approximate ergodicity, in that the variability among the tuning curves of neurons with topographic location $x_i \approx 0$ approximates variability across different $\vz$ for neuron $i_*$. This ergodicity may in principle be exploited for more efficient training and testing of the model.}

In many cortical areas the statistics of connectivity, such as connection probability and average strength, depend on several factors, including the types of the pre- and post-synaptic neurons,  the physical distance between them, or the difference between their preferred stimulus features \cite{Perin:2011,Ko:2011}.
A common statistical ensemble adopted in theoretical neuroscience posits a binary distribution for the synaptic weight $W_{ij}$: the random variable $W_{ij}$ is either zero  or, with probability $p_{ij}$, equal to a value $J_{\alpha(i)\alpha(j)}$ 
The fixed non-random  $J_{ab}$'s set the strength and signs of existing connections between neurons of different types. The connection probability $p_{ij}$ can depend on the distance between neurons $i$ and $j$ as well as on their types. In this ensemble, the $W_{ij}$ are, however,  not differentiable functions of parameters.
We instead chose a statistical ensemble for $\vc{W}$ that made the random variables, $W_{ij}$, differentiable functions of learned model parameters. For simplicity, we assumed that $W_{ij}$'s are independent and of uniform distributions with mean and range that depend on the pre- and post-synaptic types and fall off with the distance between the pre- and post-synaptic neurons over characteristic length-scales. More precisely we chose this fall-off to be gaussian, and set
\bea\label{Wensemble}
&& W_{ij} = ( J_{ab} + z_{ij}\, \delta J_{ab})\, e^{-\frac{(x_i-x_j)^2}{2 \sigma_{ab}^2}}
\qquad\qquad
a = \alpha(i), b = \alpha(j)
\\
&& z_{ij} \overset{iid}{\sim} U[0,1].
\eea
Thus $\langle W_{ij}\rangle = \bar J_{ab} e^{-\frac{(x_i-x_j)^2}{2 \sigma_{ab}^2}}$ where $\bar J_{ab} = J_{ab} + \delta\, J_{ab}/2$, while $\mathrm{Var}[W_{ij}] = \frac{1}{12} \delta J_{ab}^2\, e^{-\frac{(x_i-x_j)^2}{ \sigma_{ab}^2}}$.

Our connectivity ensemble (and hence our SSN generator) is thus characterized by 12 parameters (enumerated in Table \ref{SSNparams}), namely the elements of three $2\times 2$ matrices: the connection strength parameters $J_{ab}$, $\delta J_{ab}$, and characteristic length scales $\sigma_{ab}$, where $a,b \in \{E,I\}$. All these parameters satisfy sign constraints: $\sigma_{ab}>0$,  $J_{aE},\delta J_{aE}\geq 0$, and $J_{aI},\delta J_{aI}\leq 0$. The latter two sets of constraints ensure that any realization of $W_{ij}$ satisfies  Dale's principle \cite{Dale:1935,StrataHarvey:1999}. To enforce this constraint, the parameters we fit (those used to compute gradients) are the logarithm of the positive-definite parameters listed here.  To be explicit, the set of 12 learned parameters is \(\vth = (\log |J_{ab}|,\, \log \delta |J_{ab}|,\, \log \sigma_{ab})_{a,b \in \{E,I\}}\). Finally, a given realization of our SSN model  is specified (through its connectivity matrix) by the values of the $(2N)^2$ independent, standard uniform random variables $\vz = (z_{ij})_{i,j=1}^{2N}$. To make the dependence of the connectivity matrix on $\vz$ explicit, we denote it by $\vc{W}^{\vz}$ below.

\begin{table}
\begin{center}
\begin{tabular}{c|c}
Parameter&Description\\
\hline
$\sigma_{ab}$& connection length scales\\
$J_{ab}$& lower bounds of connection strengths \\
$\delta J_{ab}$& widths of connection strength distribution\\
\end{tabular}
\caption{The parameters of the SSN to be estimated (\(a, b \in \{E, I\}\)).}
\label{SSNparams}
\end{center}
\end{table}

Given a dataset of tuning curves, we would like to find a set of parameters, $\vth$, that produces a matching model distribution of tuning curves. {In this paper, as a proof of principle. we used simulated data, by  using the same SSN model, with parameters $\vth_0$,  to generate size tuning curves used as the training set in the GAN training}. 
During training, according to algorithm \ref{alg-1}, every time $\G(\vz)$ was evaluated, we simulated the SSN until convergence to a fixed point. The gradient $\nabla_{\vth}\G_{\vth}(\vz) = (\nabla_{\vth}G_{\vth}(s,\vz))_{s=1}^S$, is obtained from the $i_*$ components of the fixed point gradients $\partial \hat{\vr} / \partial \theta$. The expression for the latter gradient can be derived from the fixed point equation,
$\hat{\mathbf{r}} = {f} \left(\mathbf{W}^{\vz}\, \hat{\mathbf{r}} + \mathbf{I}(s)\right)$,  using the implicit derivative formula, to yield
\begin{equation}
\label{SSNgrad}
 \frac{d\hat{\mathbf{r}}}{d\vth} = (\mathbf{1} - \Phi \mathbf{W}^{\vz})^{-1} \Phi \frac{\partial \mathbf{W}^{\vz}}{\partial \vth} \hat{\mathbf{r}},
\end{equation}
where $\Phi$ is the diagonal matrix of neural gains defined by 
$
\Phi = \text{Diag}[f'\! \left(\mathbf{W}^{\vz}\,\hat{\mathbf{r}} + \mathbf{I}(s)\right)]
$.

\subsubsubsection{\textbf{Enforcing stability}}

During GAN training for fitting network parameters,  the SSN may be pushed to  parameter regions in which the network is unstable for some realizations of the randomness inputs \(\mathbf{z}\). Since our framework relies on convergence to a fixed point, and diverging rates are not biological anyway, we avoid the diverging solution by ``clipping'' the input-output nonlinearity \(f(\cdot)\) and redefining it to be
\begin{align}\label{modifpowlaw}
  f(u) =
  \begin{cases}
    k {(u)_+}^n & \text{if } u \leq V_0 \\
    r_0 \left( 1 + \frac{r_1 - r_0}{r_0} \tanh \left( n \frac{r_0}{r_1 - r_0} \frac{u - V_0}{V_0} \right) \right) & \text{if } u > V_0
  \end{cases}
\end{align}
where \(r_0 = 200\) Hz,  \(V_0 = (r_0/k)^{\frac{1}{n}}\), and \(r_1 = 1000\) Hz (this makes the nonliearity concave above the 200 Hz output level, and saturated at 1000 Hz). Note that as long as the  generator (SSN) parameters obtained by our fit have fixed points in which all rates are less than $r_0$, the model is indistinguishable from that in which the input-output nonlinearity were not modified. The modification \req{modifpowlaw} can thus allow the SSN model to venture through unstable and biologically implausible parameter regions during training and yet end up in regions with biologically plausible rates.

Even with the modified input-output relationship \req{modifpowlaw}, however, the SSN may have a limit cycle or a chaotic attractor. We therefore rejected the parameter points in which the SSN did not converge in 10 seconds of simulated time. To encourage the avoidance of such problematic parameter regions, as well as parameter regions that produce unrealistically large firing rates, we added the differentiable penalty term
\begin{align}
  \label{eq:ssn_penalty}
  \text{Penalty}_{\mathcal{G}} = 5000 \sum_{i=1}^{2N} (r_{i} - 150)_+
\end{align}
to the loss function for the generator.

\subsection{Discriminator networks}

The most studied application of GANs is in producing highly structured, high-dimensional output such as images, videos, and audio. In these applications it is beneficial to use highly complex discriminators such as deep convolutional networks. It has also been noted that the discriminator network should be sufficiently powerful so that it is capable of fully representing the data and model distributions \cite{DBLP:journals/corr/Goodfellow17,arora_gans_2017,arora_generalization_2017}. In our application, the outputs of the generator are comparatively lower dimensional objects, with less complex distributions. Correspondingly we used relatively small discriminator networks in order to speed up training time.\footnote{However, note that a $\D$ that is too simple can preclude the fitting of important aspects of the distribution. For example a linear $\D$ used in the WGAN appraoch would result in a fit that matches only the average tuning curve between model and data, and ignore tuning curve variability altogether.}

For the discriminator network, $\D$, we used  dense feedforward neural networks with two hidden layers of 128 rectified linear units each, and an output layer with a single linear readout unit. The discriminator networks used in the M1 model and SSN model fits were the same, except for the dimensionality of their input layers. Feedforward weights were initialized using the uniform distribution between \(\pm 0.0096\) and biases were initialized using a Gaussian distribution with zero mean and standard deviation 0.01. We do not use any kind of normalization or parameter regularization other than the WGAN penalty term in equation \ref{eq:loss-d}. 
We also did not do a substantial search of hyper-parameters for the discriminator networks and found that our default options  nevertheless performed quite well. This suggests that in our application, the WGAN  is relatively insensitive to the detailed structure of the discriminator as long as it is sufficiently complex.
%

\section{Experiments}
\label{sec-3}
\subsection{Complex M1 Tuning Curves}
\label{sec:lalazar-results}

We fit the feedforward network model introduced in Sec.~\ref{sec:lalazar}  using the GAN framework to trial-averaged firing rate data from monkey M1 collected by  \cite{Lalazar:2016} and  available online. The data includes neural responses for trials in which a monkey held its hand, in pronation or supinationm, in multiple target locations, as described in Sec.~\ref{sec:lalazar}. We only used the pronation conditions and pruned the data to include only neurons ($ n = 293$) which were measured in all $27$ target locations in the pronation condition. We further simplified the dataset by removing spatial scale information in the positions of target locations, which varied slightly between experimental sessions.

Reference \cite{Lalazar:2016} fit their model  by matching two specific scalar statistics (which were of interest for theoretical reasons)  characterizing tuning curves. We will instead adopt a more agnostic approach and let the WGAN fit the full multi-dimensional tuning curve distribution.
Moreover,  in \cite{Lalazar:2016}, the model and data tuning curves were individually normalized so that their shape, but not their overall rate scale, was fit to data. As described in Sec.~\ref{sec:lalazar}, we instead included a scaling parameter, $J$, for the feedforward connection strengths and found that with this addition the model, without normalization, is actually able to account for the variability in average firing rate across neurons as well. 

\begin{figure}
\includegraphics[width = 5in]{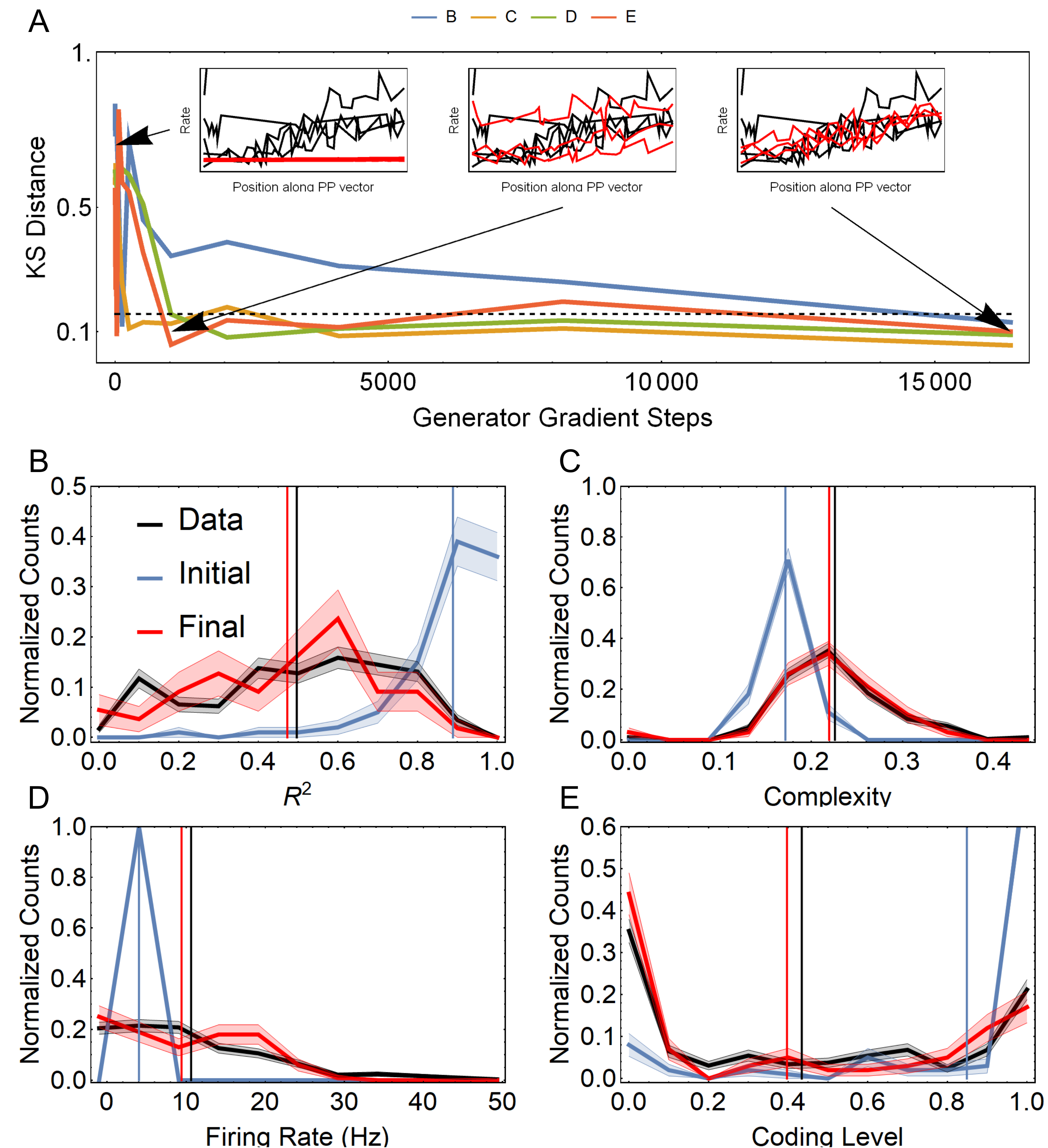}
\caption{Summary of the feedforward model fit to the M1 tuning curve data of Ref.~\cite{Lalazar:2016}. \textbf{A}: Kolmogorov-Smirnov (KS) distance during training between the data and model distributions, shown in panels B-E, of four summary statistics characterizing the tuning curves. The dashed line shows the KS distance corresponding to significance level $P < 0.05$ between samples of data ($n = 293$) and simulated ($n = 100$) tuning curves. The insets show data and model tuning curves at different points throughout training; the plotted tuning curves are the projections of the 3D tuning curves along the preferred position (PP) vector which is obtained by a linear fit to the 3D tuning curve (see \cite{Lalazar:2016}). \textbf{B-E}: histograms of four tuning curve statistics ($R^{2}$, complexity score, rates, and coding level) showing the data distribution (black), the initial model distribution (blue) and the final model distribution after fitting (red). Vertical lines show the mean value of each histogram.}
\label{FFfigure}
\end{figure}

The results of our fitting procedure are summarized in Fig.~\ref{FFfigure}. We show the initial and final histograms for four statistics characterizing the tuning curves: $R^{2}$ of the linear fit to the tuning curve, complexity score, firing rate (in all conditions),  and coding level, \ie~the fraction of conditions with rate significantly greater than zero (which we took to mean larger than 5 Hz). Ref. \cite{Lalazar:2016} defined the complexity score of a tuning curve to be the standard deviation of the absolute response differences  
 between nearest neighbor locations on the $3\times 3\times 3$ lattice of hand positions (the tuning curve was first normalized so that its responses ranged from $-1$ to $1$). 
The first two properties, $R^2$ and complexity score, are the two statistics used by \cite{Lalazar:2016} to fit their model; they provide two measures for the degree of nonlinearity of the tuning curve.  

We measured the mismatch between the model and data distributions using the Kolmogorov-Smirnov (KS) distance. 
Figure \ref{FFfigure}A shows the KS distance between the model and data distributions for the chosen four properties throughout model training using stochastic gradient descent, with the dashed line showing $p < 0.05$. The distribution of complexity (Fig.~\ref{FFfigure}C) is very accurately captured. It is also notable that the trained model fits the distribution of firing rates and coding levels (Fig.~\ref{FFfigure}D-E) quite well: in the work of Ref.~\cite{Lalazar:2016} the scale of rates was not fit at all (tuning curves were individually normalized) and the fit of coding levels was enforced by hand, as described in Sec.~\ref{sec:lalazar} and below.

 The relatively poorer performance in capturing $R^2$ statistics (Fig.~\ref{FFfigure}B), compared to the enhanced model of \cite{Lalazar:2016}, may be due to the more flexible way by which they fit individual neuron thresholds (which we did not do). The fact that our distribution of neuron coding levels  matches the data quite well suggests that the marginal variability in threshold was nevertheless learned properly by our model. Thresholds in the original work were, however, not independently sampled but were set after other random inputs, $\vz$, so as to match the response coding level of the model neuron to that of a randomly sampled data neuron. Because of this approach, the distribution of thresholds in \cite{Lalazar:2016}'s enhanced model was conditioned upon the other random model inputs, $\vz$. If these conditional dependencies are important for fitting the $R^{2}$ distribution, the simpler model we implement here is not sufficiently expressive to capture the $R^{2}$ distribution in the data.

\subsection{Stabilized Supralinear Network}

To demonstrate the effectiveness of our proposed approach in its application to complex, dynamical models, we used the GAN framework to train the connectivity parameters of the SSN model described in Sec.~\ref{subsec:SSNmodel}. We fit this model to simulated data: $1000$ tuning curves generated from an SSN with the connectivity parameters listed in table \ref{tab:ssn_true_params}.
This set of simulated tuning curves serves as the empirical data distribution for this GAN fit.
We fixed all other parameters of both the true and trained SSN and their stimulus inputs as follows:
$N=201$, $k=0.01$, $n=2.2$, $\tau_E = 16$ ms, $\tau_I = 2$ ms, $A = 20$, $l = 1/4$, $(b_s)_{s=1}^S =(0.5, 1, 2, 4, 6) $ (for testing purposes, after fitting, we generated tuning curves from the trained SSN using a larger set of stimulus sizes $b$).
To train the learned SSN, we initialized the network parameters \((\sigma_{ab}, J_{ab}, \delta J_{ab})_{a,b \in \{E,I\}}\)
to 0.37 times the values of the true parameter. This was done to assure that the SSN began training in a stable regime (see Sec.~\ref{subsec-42} below) and started with parameters that were sufficiently far away from the true parameters so as to make the training non-trivial.

\begin{table}[h]
    \centering
    \begin{tabular}{c|c|c|c|c}
        \text{Parameter}&EE&EI&IE&II\\
        \hline
        \(J_{ab}\) & 0.0957 & .0638 & .1197 & .0479\\
        \(\delta J_{ab}\) & $.7660$ & $.5106$ & $.9575$ & $.3830$\\
        \(\sigma_{ab}\) & .6667 & .2 & 1.333 & .2 \\
    \end{tabular}
    \caption{The parameter values used for generating the trainig set tuning curves. The columns correspond to different $(a,b)$ possiblities.}
    \label{tab:ssn_true_params}
\end{table}

To quantify the goodness-of-fit between the model and data distributions of tuning curves we compare the distribution of four attributes or statistics characterizing the tuning curves: the suppression index, preferred stimulus size, maximum firing rate, and peak width, defined as follows.
The suppression index for a neuron (or tuning curve) measures how much the response of that neuron is suppressed  below its peak level at large stimulus sizes. It is defined by
\begin{align*}
    \text{Suppression Index} = 1 - \frac{\hat r(\max  (b))}{\max_b (\hat r(b))}
\end{align*}
where \(\hat r_{i_*}(b)\) is the steady-state response of the neuron to the stimulus with size \(b\).
The maximum firing rate is simply \(\max_{b} \hat r(b)\), and the preferred size is \(\operatorname{arg}\max_{b} \hat r(b)\).
Finally, the peak width is a variant of the inverse participation ratio \cite{Mirlin1999} that has the correct units, and limiting behavior, to be interpreted as the width of the tuning curve peak, and is defined by
\begin{align}
\label{peak_width}
\text{Peak Width} = \left(\sum_{b}\left(\frac{\hat r(b)}{\sum_{b'}\hat r(b')}\right)^{2}\right)^{-1}.
\end{align}

\begin{figure}
\centering
\includegraphics*[width=5in]{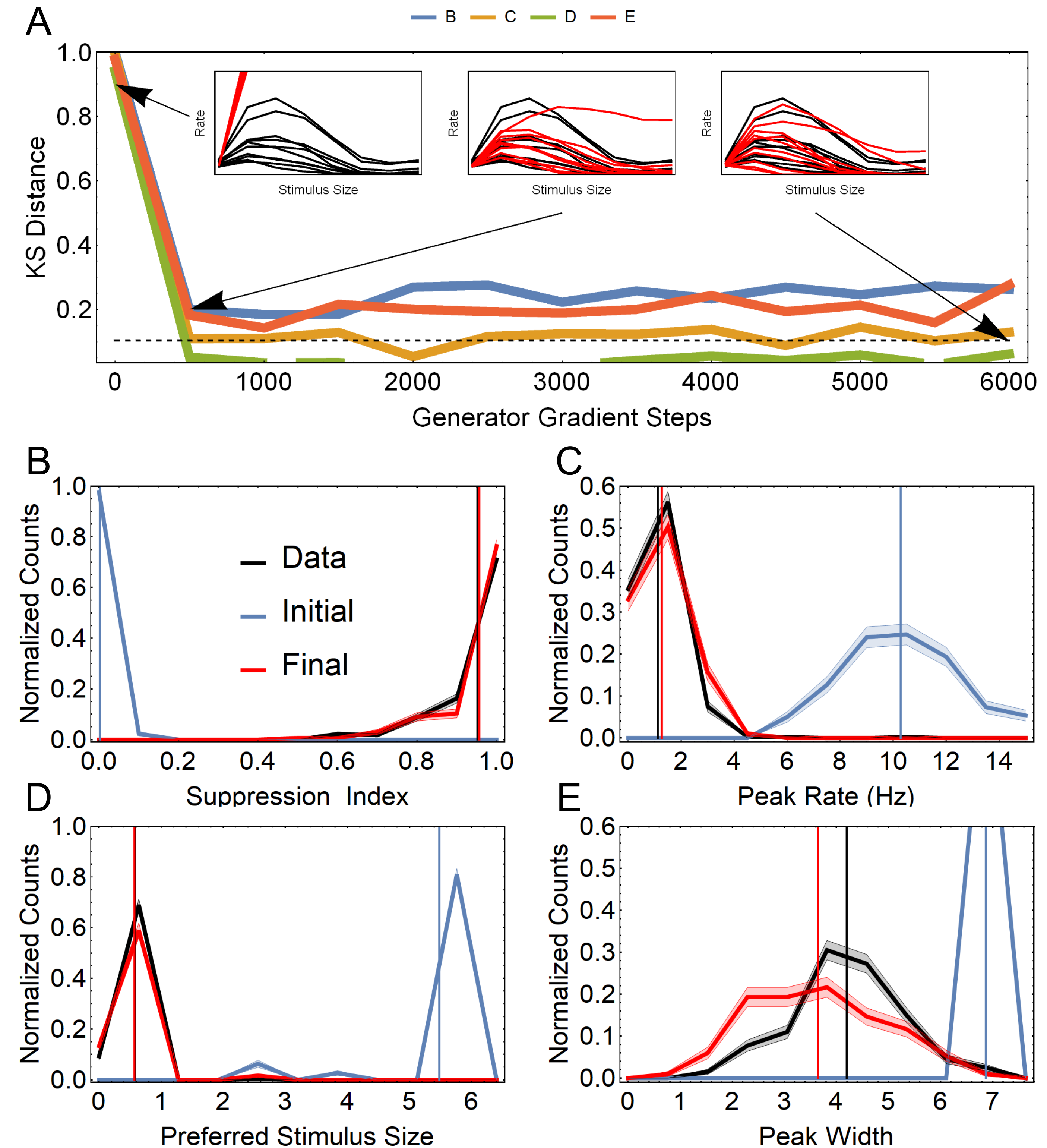}
\caption{Summary of the recurrent SSN model fit to simulated V1 tuning curve data. \textbf{A}: KS distance during training between the data and model distributions, shown in panels B-E, of four summary statistics characterizing the tuning curves. The dashed line shows the KS distance corresponding to significance level $P < 0.05$ between a sample of data ($n = 400$) and simulated ($n = 300$) tuning curves. The insets show data and model tuning curves at different points throughout training. \textbf{B-E}: histograms of four tuning curve statistics (suppression index, peak rate, preferred size, and peak width) showing the data distribution (black), the initial model distribution (blue) and the final model distribution after fitting (red). Vertical lines show the mean value of each histogram.}
\label{fig:ssn}
\end{figure}

Figures \ref{fig:ssn} B-E provide comparisons of the initial and final distributions of these tuning curve attributes.
As above, we measured the mismatch of these distributions under the trained and true SSN using the KS distance. The KS distance for all distributions decreased during learning, indicating that the fit of the model distributions to the data were in fact improving, and the final fit of all distributions is very good (Figure \ref{fig:ssn}A). 
\footnote{Note that the KS-test is very stringent: in the limit where the number of samples grows large even quite similar distributions will be reported as significantly different under this measure, however KS distance alone provides a convenient way to quantify improvement during training, even though the associated p-value may be misleading.}

\section{Possible issues}
\label{sec-4}

\subsection{Optimization difficulties}

As with any gradient based fit it is possible for a GAN or WGAN to become stuck in sub-optimal local minima for the generator. Furthermore, it is an open question whether GAN training will always converge \cite{DBLP:journals/corr/Goodfellow17,nagarajan_gradient_2017,heusel_gans_2017}. As research in GANs and non-convex optimization advances this issue will be improved. For now avoiding this pitfall will be a matter of the user judging the quality of fit after the fit has converged. Starting the gradient descent with several different initial conditions for $\vtheta$ (parameters of $\G$) can also help, with some initializations leading to better final fits.

\subsection{Dynamical stability in recurrent generative models}
\label{subsec-42}
 
 When the generator is a recurrent neural network (RNN) with its output based on a  fixed point of the RNN (as was the case in our SSN experiment), an important potential issue is lack of convergence to a stable fixed point for some choices of recurrent network parameters. 
In the experiment with SSN we initialized the generator network in such a way that it had a stable fixed point for almost all realizations of \(\mathbf{z}\). For the SSN this would generically be the case when recurrent excitation (which has destabilizing effects) is sufficiently weak. Hence initializations with small \(J_{EE}\) and \(\delta J_{EE}\) are good choices. In addition, a relatively large SSN size \(N\) improves the stability issue because random relative fluctuations in total  incoming synaptic weights are small when the number of post-synaptic neurons is large. Thus, when $N$ is large, for a given choice of network parameters, $\theta$, either the network converges to a stable fixed point for almost all $\vz$, or almost never does. Finally, to avoid entering  parameter regions leading to instability during training, we used the tricks introduced at the end of Sec.~\ref{subsec:SSNmodel}.

The issue of fixed point stability is reminiscent of the recurrent weight initialization problem in learning of RNNs. In a non-back-propagation based paradigm called the Echo State Networks (ESN) \cite{Jaeger2004}, the recurrent weights are configured in such a way that the network dynamics always settle in a fixed point with any stationary input. Jaeger \cite{Jaeger2004} showed that this can be achieved by tuning the spectral radius of the recurrent weight matrix to be smaller than 1 for the $\tanh(x)$ nonlinearity. This helps the network to continuously flush out the memory of old inputs and become ready to store new incoming signals. Although the recurrent weights are not updated in the original ESN formalism, Sutskever et al.~\cite{sutskever_importance_2013} applied this idea also to the initialization of  recurrent weights learned using the back-propagation through time (BPTT) algorithm. Since an overly small spectral radius is also problematic for the BPTT algorithm due to vanishing gradient \cite{pascanu_difficulty_2013}, they found that choosing the spectral radius to be just above instability (specifically, 1.1) achieves a good balance between stability and speed of training. The spectral radius slightly less than 1 (0.95) also has been used \cite{pascanu_difficulty_2013}. Unlike those RNNs which learn general dynamics, we do not have the vanishing gradient problem in our SSN example, because in that case the spectral radius is less than 1 with the gradient calculated without BPTT, using \req{SSNgrad}. Thus, we are led to believe that starting learning from a very stable dynamical regime, which is analogous to a small spectral radius, is useful in general when the output of the generator is the fixed point of a dynamical system.

\subsection{Unidentifiability of parameters}

In fitting generative models it is possible for models with widely different  parameters to result in nearly identical output distributions. In our case, this corresponds to cases in which   networks with widely divergent connectivity or single-cell parameters  nevertheless generate very similar tuning curve distributions. In such cases it would not be possible to use our approach to make precise inferences about network parameters (\eg connectivity parameters)  using tuning curve data.  
In the case of our SSN experiment, for example, we trained the generative model using only tuning curves with respect to stimulus size (measured at a few different sizes). In fact, there is no general reason why a low-dimensional output of the network, such as size tuning curves, would provide sufficient information for constraining all model parameters. 
Even in such cases, however, a richer dataset of tuning curves can allow for further constraining of model parameters. 
 Fortunately there is nothing in our approach that would prevent one from using joint tuning curves with respect to several stimulus parameters (\eg joint tuning curves with respect to both stimulus size and stimulus intensity). 
 Generally, it should be the case that the more high-dimensional the tuning curves used in training, \ie the larger the number of stimulus parameters and conditions they span, the more identifiable the network parameters. Thus with sufficiently rich datasets, our GAN-based method provides a promising way to infer biophysical networks parameters, such as those governing connectivity statistics.  
 Furthermore, our approach can in principle be used to \emph{design} experiments, i.e., optimally choose the stimulus conditions and quantities to be recorded, to maximize the identifiability of the parameters of a given model.

\section{Conclusion}
\label{sec-5}

Developing theoretically grounded models that can capture the diversity and heterogeneity of neural response selectivities is an important task in computational neuroscience. The ability to fit such models directly to neural data, \eg~to datasets of heterogeneous tuning curves from some brain area, can greatly facilitate this pursuit. However, the statistical fitting of biologically motivated mechanistic models of brain circuitry  to neural data, using likelihood-based approaches, is  often intractable. A practical solution used in the past has been to instead use  models that are designed primarily to have tractable likelihoods and to be easy to fit. The elements and parameters of such models may not have mechanistic, biological interpretations. 

Here, we demonstrated that Generative Adversarial Networks enable the fitting of theoretically grounded models to the full distribution of neural data. Conceptually, we view  mechanistic network models with randomness in their structure or biophysical parameters as  generative models for diverse tuning curves.
In Sec.~\ref{sec-2} we reviewed the basic GAN setup and the  WGAN algorithm, as developed in the machine learning literature. We used this method to successfully fit  two representative example models from theoretical neuroscience to tuning curve data, demonstrating that this technique is applicable to a wide range of network models including feedforward  and recurrent models of cortical networks. In Sec.~\ref{sec-4}, we reviewed some of the potential pitfalls of our approach in an attempt to make the path clearer for other  applications of this approach.

There are several directions in which the current work can be extended. For example, the methodology can be generalized such that it is able to handle cases of datasets with missing data; it is often the case in experiments that not all neurons are recorded in all stimulus conditions.  As another extension, similar methodology can be used to fit mechanistic recurrent network models to simultaneously recorded neural activity, to capture, \eg~the distribution of noise correlations between pairs of neurons in some cortical area. 
 More generally, the potential for the application of newly developed methods for fitting implicit generative models in neuroscience is broad.

\subsubsection*{Acknowledgments}
The authors thank Daniel Lowd and Caleb Holt for inspiring conversations.

\small{
\bibliographystyle{plos}
\bibliography{GANbib.bib}
}

\end{document}